\documentstyle[preprint,prc,aps,epsfig]{revtex}
\begin{document}
\draft

\title{Modelling nucleon-nucleon scattering above 1 GeV}

\author{K. O. Eyser,$^{1,2,}$\thanks{Electronic address: oleg@kaa.desy.de}
        R. Machleidt,$^{1,}$\thanks{Electronic address: machleid@uidaho.edu}
        W. Scobel,$^{2,}$\thanks{Electronic address: scobel@kaa.desy.de}
        and the EDDA Collaboration\thanks{see reference [1]}}

\address{
$^1$Department of Physics, University of Idaho, Moscow, Idaho 83844, USA
\\
$^2$ Institut f\"ur Experimentalphysik, Universit\"at Hamburg, 
D-22761 Hamburg, Germany
}

\date{\today}

\maketitle

\begin{abstract}
Motivated by the recent measurement of proton-proton
spin-correlation parameters up to 2.5 GeV laboratory energy, 
we investigate models for nucleon-nucleon ($NN$) scattering above 1 GeV.
Signatures for a gradual failure of the traditional meson model
with increasing energy can be clearly identified.
Since spin effects are large up to tens of GeV,
perturbative QCD cannot be invoked to fix the problems.
We discuss various theoretical scenarios and come to the conclusion
that we do not have a clear phenomenological understanding of the spin-dependence
of the $NN$ interaction above 1 GeV.
\end{abstract}


\section{Introduction}

The force between two nucleons has been studied for many decades.
Based upon the Yukawa idea~\cite{Yuk35}, meson theories were developed
in the 1950s~\cite{TMO52,BW53,Gar55,SM58} and 60s~\cite{OBEP67,note1}.
However when, in the 1970s, quantum chromodynamics (QCD) emerged
as the generally accepted theory of strong interactions, those ``meson theories''
were demoted to models and the attempts to derive the nuclear force
in fundamental terms had to start all over again.

The problem with a derivation from QCD is that this theory is
nonperturbative in the low-energy regime
characteristic for nuclear physics and direct solutions
are impossible. Therefore,
QCD inspired quark models were fashionable for a 
while---in the 1980s~\cite{MW88}.
However, since they are---admittedly---just another set of models,
they do not represent any progress in fundamental terms.
If one has to resort to models anyhow,
then one can, as well, continue to use meson models:
they are relatively easy to build, the predictions are 
quite quantitative, and the underlying picture is very intuitive:
mesons of increasing masses are exchanged, creating contributions
of decreasing ranges until the range is sufficiently short such
that it may be considered irrelevant for nuclear physics purposes.

A certain breakthrough occured, when the effective fieldtheory (EFT)
concept was introduced and applied to low-energy QCD~\cite{Wei79}. 
Based upon
these ideas, Weinberg showed in 1990~\cite{Wei90}, that a systematic
expansion of the nucleon-nucleon ($NN$) amplitude exists in terms
of $(Q/\Lambda_\chi)^\nu$, where $Q$ denotes a generic nucleon momentum,
$\Lambda_\chi \approx 1$ GeV is the chiral symmetry breaking scale,
and $\nu \geq 0$.
This is known as chiral perturbation theory ($\chi$PT) which is
equivalent to low-energy QCD.

Weinberg's initial work~\cite{Wei90} created a lot of interest and
activity~\cite{ORK94,Kol99,CPS92}
that has been going on now for more than a 
decade~\cite{KBW97,EGM98,Kai99,BK02}.
As a result, we have
today a rather precise understanding of the nuclear force in terms
of $\chi$PT~\cite{EM02,EM02a,EM03}. 
However, the energy range appropriate for $\chi$PT is
very limited; afterall, 
$\chi$PT is a low-momentum expansion, good only for momenta
$Q\ll \Lambda_\chi \approx 1$ GeV.
The most advanced calculations to date go to fourth order~\cite{EM02a,EM03}
at which $NN$ scattering 
can be described satisfactorily
up to lab.\ energies ($T_{\rm lab}$) of about 300 MeV.
For higher energies, more orders must be included.
However, since the number of terms increases dramatically with each higher
order (cf.\ Ref.~\cite{EM02a}), $\chi$PT will be unpractical
and unmanagable around order five or six.
It is, thus, safe to state that $\chi$PT is limited to
$T_{\rm lab} < 0.5$ GeV.

There is, of course, a need to understand $NN$ scattering 
also above 0.5 GeV.
Naively one might expect that
perturbative QCD (pQCD) should be useable
above the scale of the low-energy EFT.
Unfortunately, this is not true.
Energies of the order of 1 GeV are far too low to invoke
pQCD. Thus, in the energy range that stretches from
about 0.5 GeV to, probably, tens of GeV, we are faced with the dilemma that
we have presently no calculable theory at our disposal.
In principal, it should be possible to
apply lattice QCD in this energy regime.
However, such
calculations are not available, at this time. They are
an interesting prospect for the future.

For theoretical physics, it is not uncommon to encounter
such problems.
Typically, the preliminary way out is to build models.
The hope is that reasonably constructed models may provide
insight which may ultimately lead to
a solution on more fundamental grounds.
The `standard model' for the nuclear force is relativistic meson-exchange.
In the past, meson models have been constructed
and shown to describe $NN$ scattering up to about 1 GeV 
satisfactorily~\cite{Mac89,KS80,DKS82,Lom82,Lee83,FT84,HM87,Els88a,Els88b,FL90,SM90}.

However, above 1 GeV, there remains a large energy range
where pQCD is still not applicable. Traditionally, this energy region has
been the stepchild of the theoretical profession.
Recently, a large number of precise $pp$ scattering
data up to 2.8 GeV have been 
measured~\cite{EDDA1,EDDA2,EDDA3,All98,All99,All01}.
The obvious question is: Do we understand 
these data, their angular and energy dependence? The focus of this paper will 
be on the spin observables that are more exclusive than the spin averaged 
cross sections.

This paper is organized as follows.
In Sec.~II, we present a typical model that is known to be 
appropriate for energies up to about 1 GeV.
In Sec.~III, this model is applied for energies above 1 GeV,
and some modifications necessary for those higher energies are
introduced. Sec.~IV is then devoted to spin observables.
We conclude the paper with Sec.~V, where we elaborate on the unsolved problems
of the energy region under consideration.

\section{Relativistic meson-exchange model for $NN$ scattering
at intermediate energies}

The simplest meson model for the nuclear force
is the so-called one-boson-exchange (OBE) model which
takes only single-particle exchanges into account~\cite{note2}.
Typically, the mesons with masses below the nucleon
mass are included.
Most important, are the following four mesons:
\begin{itemize}
\item The pseudoscalar pion with a mass of about 138 MeV.
It is the lightest meson and provides the long-range part of the potential 
and most of the tensor force.
\item The $\rho$ meson, a 2$\pi$ $P$-wave resonance of about 770 MeV.
Its major effect is to cut down the 
tensor force provided by the pion at short range---to a realistic size.
\item The $\omega$ meson, a 3$\pi$ resonance of 783 MeV and spin 1.
It creates a strong repulsive central 
force of short range (`repulsive core') and most of the nuclear spin-orbit force.
\item The $\sigma$ boson of about 550 MeV. It
 provides the intermediate 
range attraction necessary for nuclear binding and  can 
 be understood 
as a simulation of the correlated $S$-wave 2$\pi$-exchange.
\end{itemize}
Besides these four bosons, we include also the $\eta(547)$,
which brings the total number to five.
The quantum numbers characterizing these mesons
(like, spin, parity, isospin) are shown in Table~I.

The following Lagrangians describe the coupling of these
mesons to nucleons:
\begin{eqnarray}
{\cal L}_{pv}&=& -\frac{f_{ps}}{m_{ps}}\bar{\psi}
\gamma^{5}\gamma^{\mu}\psi\partial_{\mu}\varphi^{(ps)}
\label{eq_pv}
\\
{\cal L}_{s}&=& -g_{s}\bar{\psi}\psi\varphi^{(s)}
\label{eq_s}
\\
{\cal L}_{v}&=&-g_{v}\bar{\psi}\gamma^{\mu}\psi\varphi^{(v)}_{\mu}
-\frac{f_{v}}{4M} \bar{\psi}\sigma^{\mu\nu}\psi(\partial_{\mu}
\varphi_{\nu}^{(v)}
-\partial_{\nu}\varphi_{\mu}^{(v)})
\label{eq_v}
\end{eqnarray}
where $M$ is the nucleon mass and $m_\alpha$ a meson mass.
$\psi$ denotes the nucleon and $\varphi^{(\alpha)}_{(\mu)}$ the meson fields
(notation and conventions as in Ref.~\cite{BD64}).
For isospin-1 (isovector) mesons, $\varphi^{(\alpha)}$ is to be replaced
by {\boldmath $\tau \cdot \varphi^{(\alpha)}$}
with $\tau^{i}$ ($i=1,2,3$) the usual Pauli matrices.
$ps$, $pv$, $s$, and $v$ denote pseudoscalar, pseudovector, scalar, and vector
couplings/fields, respectively.
For the pseudoscalar mesons $\pi$ and $\eta$, we use the pseudovector 
coupling, Eq.~(\ref{eq_pv}), as suggested by chiral symmetry.
The scalar boson $\sigma$ couples via the scalar Lagrangian,
Eq.~(\ref{eq_s}), and
the vector mesons $\rho$ and $\omega$ interact through the
Lagrangian Eq.~(\ref{eq_v}).
The coupling constants, $g_\alpha$ and $f_\alpha$, are given in Table~I in terms of 
$g^2_\alpha/4\pi$ and
$f^2_\alpha/4\pi$, respectively.

Based upon the above Langrangians, the 
one-particle-exchange Feynman diagrams (Fig.~1)
can be evaluated straightforwardly (see Ref.~\cite{Mac89}
for details).
The OBE potential is then defined as the sum of the 
Feynman amplitudes created from the five mesons:
\begin{equation}
\bar{V}=\sum_{\alpha=\pi,\eta,\sigma,\rho,\omega}
\bar{V}_{\alpha}
\label{eq_obep}
\end{equation}
Explicit expressions for the Feynman amplitudes, 
$\bar{V}_{\alpha}$, can be found in Refs.~\cite{Mac89,Mac93}.
We note that we modify these Feynman amplitudes by
applying, at each meson-nucleon vertex,
a form factor which has the analytical form
\begin{equation}
F_\alpha[(\vec{q'}-{\vec q})^2]=
\left( \frac{\Lambda^2_\alpha-m^2_\alpha}
{\Lambda^2_\alpha+(\vec{q'}-{\vec q})^2} 
\right)^{n_\alpha}
\,,
\label{eq_ff}
\end{equation}
where $\vec{q}$ and $\vec{q'}$ denote the nucleon momenta
in the center-of-mass (c.m.)\ frame in the initial and final state, respectively;
and $(\vec{q'}-\vec{q})$ is the momentum transfer
between the two interacting nucleons.
$\Lambda_\alpha$ is called the cutoff mass.
We use $n_\alpha =1$ for all vertices with the exception of the
$N\Delta\rho$ vertex where $n_\alpha=2$ is applied (s.\ below).
The form factor suppresses the contributions from high momentum
transfer which is equivalent to short distances. 
This is necessary to make loop integrals (and the solution
of the Lippmann-Schwinger equation) convergent
and suggested by the extended (quark) substructure of hadrons.

For the energies to be considered in this study,
it is mandatory to use a relativistic formalism.
Relativistic $NN$ scattering is described by the
Bethe-Salpeter equation~\cite{SB51}.
Unfortunately, this four-dimensional equation is difficult
to solve. Therefore, so-called three-dimensional
reductions have been proposed which are more amenable
to numerical solution. We will use the relativistic 
three-dimensional Thompson equation~\cite{Tho70} which reads,
\begin{equation}
\bar{T}(\vec{q'},\vec{q};\sqrt{s})=\bar{V}(\vec{q'},\vec{q})+
\int d^3k\:
\bar{V}(\vec{q'},\vec{k})\:
\frac{M^{2}}{E_{k}^{2}}\:
\frac{1}
{\sqrt{s}-2E_{k}+i\epsilon}\:
\bar{T}(\vec{k},\vec{q};\sqrt{s}) \, ,
\label{eq_thom}
\end{equation}
where $\bar{T}$ denotes the invariant scattering amplitude
and $\sqrt{s}$ is the total energy in the
c.m.\ frame;
$\sqrt{s}=2E_q$ with
$E_q\equiv \sqrt{q^2+M^2}$
and $q\equiv |\vec{q}|$ the momentum
of one nucleon in the 
c.m.\ frame,
which is related to the lab.\ energy of the projectile by
$T_{\rm lab} = 2q^2/M$.
It is convenient to define
\begin{equation}
{T}(\vec{q'}, \vec{q})
 = {\frac{M}{E_{q'}}}\: \bar{T}(\vec{q'}, \vec{q})\:
 {\frac{M}{E_{q}}}
\end{equation}
and
\begin{equation} 
{V}(\vec{q'},\vec{q})
 = {\frac{M}{E_{q'}}}\:  \bar{V}(\vec{q'},\vec{q})\:
 {\frac{M}{E_{q}}}\: .
\label{eq_obep2}
\end{equation}
With this,
we can rewrite Eq.~(\ref{eq_thom}) as
\begin{equation}
{T}(\vec{q'},\vec{q};\sqrt{s})={V}(\vec{q'},\vec{q})+
\int d^3k\:
{V}(\vec{q'},\vec{k})
\frac{1}
{\sqrt{s}-2E_k+i\epsilon}
{T}(\vec{k},\vec{q};\sqrt{s})
\label{eq_thom2}
\end{equation}
which resembles a Lippmann-Schwinger equation with relativistic energies.

In the framework of the relativistic three-dimensional reduction of
the Bethe-Salpeter equation applied here (Thompson equation),
the OBE potential is real at all energies and, therefore, 
suitable only for $NN$ scattering
below the inelastic threshold. Above $T_{\rm lab} \approx 290$ MeV,
pions can be produced in $NN$ collisions. 
A model that is expected to have validity at intermediate
energies needs to take the inelasticity due to pion-production
into account.
It is well-known that, below about 1.5 GeV, pion production
proceeds mainly through the formation of the $\Delta(1232)$
isobar which is a pion-nucleon resonance with spin and isospin 3/2.
The next higher resonance is the $N^*(1440)$,
also known as Roper resonance, with spin and isospin 1/2~\cite{PDG02}. 
This resonance was included in a meson model for $NN$ scattering
up to 2 GeV constructed by Lee~\cite{Lee84} and found to contribute
less than 1 mb to the inelastic cross section even at 2 GeV.
A recent exclusive measurement of two-pion production in
$pp$ scattering at 775 MeV finds cross sections that 
can be attributed to the Roper resonance of less
than 0.1 mb~\cite{Pat03}. Thus below 2 GeV,
the $N^*(1440)$ is much less important
than the $\Delta(1232)$.
Therefore, we introduce only the $\Delta$ as 
an additional degree of freedom. Consequently, we have now,
besides the $NN$ channel, two more two-baryon channels,
namely, $N\Delta$ and $\Delta\Delta$. 

Since all channels
have baryon number two, transitions between these channels are
allowed, i.~e., the channels ``couple''. 
Mathematically this produces a system of coupled
equations for the scattering amplitudes.
In operator notation, one can write:
\begin{equation}
T_{ij} = V_{ij} + \sum_{k} V_{ik} \: g_k \: T_{kj} \, ,
\label{eq_coup}
\end{equation}
where each subscript $i, j,$ and $k$ denotes a two-baryon channel
( $NN$, $N\Delta$, or $\Delta\Delta$), and $g_k$ is the
appropriate two-baryon propagator.
In principal, there are nine transition potentials, $V_{ij}$,
which reduce to six due to time-reversal.
Three of them, namely,
$V_{N\Delta,N\Delta}$,
$V_{N\Delta,\Delta\Delta}$,
$V_{\Delta\Delta,\Delta\Delta}$,
involve $\Delta\Delta \alpha$ vertices, where $\alpha$
is a non-strange meson. Exploiting the usual symmetries,
such vertices can be constructed;
however, there is no way to test empirically if the assumptions
about these vertices are realistic.
Therefore, such constructs are
beset with large uncertainties, which is why we
omit them.
The consequence is that the system of coupled equations,
Eq.~(\ref{eq_coup}), decouples and the $T$-matrix of $NN$ scattering,
$T\equiv T_{NN,NN}$, is
the solution of just one integral equation:
\begin{equation}
T = V_{\rm eff} + V_{\rm eff} \; g_{NN} \; T \, ,
\label{eq_t}
\end{equation}
with
\begin{equation}
V_{\rm eff} = V_{NN,NN} 
+ V_{NN,N\Delta} \; g_{N\Delta} \; V_{N\Delta,NN}
+ V_{NN,\Delta\Delta} \; g_{\Delta\Delta} \; V_{\Delta\Delta,NN} \, ,
\label{eq_eff}
\end{equation}
where $V_{NN,NN}$ is the $V$ given in Eq.~(\ref{eq_obep2})
which is based upon Eq.~(\ref{eq_obep}) and shown in Fig.~1.
The last two terms on the r.h.s.\ of the above equation
are depicted in Fig.~2. 

Because of isospin conservation,
the transition potentials containing $N\Delta\alpha$ vertices
can only involve isovector mesons.
Thus, 
\begin{eqnarray}
V_{NN,N\Delta} & = & \sum_{\alpha=\pi,\rho} 
V^{\alpha}_{NN,N\Delta} \, , \\
V_{NN,\Delta\Delta} & = & \sum_{\alpha=\pi,\rho} 
V^{\alpha}_{NN,\Delta\Delta}
\, .
\end{eqnarray}
The amplitudes, 
$V^{\alpha}_{NN,N\Delta}$ and
$V^{\alpha}_{NN,\Delta\Delta}$,
with $\alpha = \pi, \rho$,
are derived from the interaction Langrangians:
\begin{eqnarray}
{\cal L}_{N\Delta\pi} & = & -\frac{f_{N\Delta\pi}}{m_{\pi}}
 \bar{\psi}{\bf T}\psi^{\mu}\partial_{\mu}
{\mbox{\boldmath $\varphi$}}^{(\pi)}
+ {\rm H.c.} \, , \\
{\cal L}_{N\Delta\rho} & = & i\frac{f_{N\Delta\rho}}{m_{\rho}}
\bar{\psi}\gamma^{5}\gamma^{\mu}{\bf T} \psi^{\nu}(\partial_{\mu}
{\mbox{\boldmath $\varphi$}}^{(\rho)}_{\nu}
- \partial_{\nu}{\mbox{\boldmath $\varphi$}}_{\mu}^{(\rho)})
+{\rm H.c.} \, ,
\end{eqnarray}
where $\psi_{\mu}$ is a Rarita-Schwinger field~\cite{RS41,Lur68,Dum83}
 describing the (spin $\frac{3}{2}$) $\Delta$-isobar
and ${\bf T}$ denotes an isospin transition operator that acts between
isospin-$\frac12$ and isospin-$\frac32$ states.
H.c.\ stands for hermitean conjugate.
The transition potentials $V^{\pi}_{NN,N\Delta}$ and
$V^{\pi}_{NN,\Delta\Delta}$ can be found in Ref.~\cite{HM77}
and $V^{\rho}_{NN,N\Delta}$ and
$V^{\rho}_{NN,\Delta\Delta}$ are given in Ref.~\cite{Hol78}.
We use these relativistic transition potentials in conjunction
with static meson propagators that take the delta-nucleon
mass difference into account. The vertices involved in the
transition potentials are multiplied with form factors
of the type Eq.~(\ref{eq_ff}).

The two-baryon propagators involved in Eqs.~(\ref{eq_t}) and (\ref{eq_eff})
are,
\begin{eqnarray}
g_{NN} & = & \frac{1}{\sqrt{s}-2E_k + i\epsilon} \,, \\
g_{N\Delta} & = & \frac{1}{\sqrt{s} - E_k - 
  \tilde{E}^\Delta_k(\sqrt{s})
      } \,, \\
g_{\Delta\Delta} & = & \frac{1}{\sqrt{s} - 2\tilde{E}^\Delta_k(\sqrt{s})} \,,
\end{eqnarray}
where
$\tilde{E}^\Delta_k(\sqrt{s}) = \sqrt{k^2 + 
\tilde{M}_\Delta^2(\sqrt{s})
}$
with
$\tilde{M}_\Delta(\sqrt{s}) = M_\Delta - i\Gamma(\sqrt{s})/2$
a complex $\Delta$-mass.
The real part of the $\Delta$ mass is the wellknown physical
mass, $M_\Delta = 1232$ MeV.
The imaginary part, which is associated with the decay-width
of the $\Delta$-isobar,
creates the inelasticity
in our model and simulates pion production.
It is calculated from the
self-energy of the $\Delta$-isobar that
is obtained from a solution of the
Dyson equation in which the
$\Delta$ is coupled virtually to
the $\pi N$ decay channel~\cite{FT84,HM87,Els88b}.
$\Gamma(\sqrt{s})$,
is energy-dependent and the threshold is
$\sqrt{s}=2M+m_\pi$ 
for diagrams with one intermediate
$\Delta$ state and
$\sqrt{s}=2M+2m_\pi$ 
for two intermediate $\Delta$.
Below these thresholds,
$\Gamma(\sqrt{s})$ vanishes. 
Note that, due to isospin conservation, $N\Delta$ diagrams contribute
only in isospin $T=1$ $NN$-states,
while $\Delta\Delta$ diagrams contribute to all states.
Consequently, in $T=0$, only 
double-$\Delta$ diagrams contribute (besides the usual OBE
contributions, Fig.~1). 
This explains the thresholds for inelasticity seen in Fig.~3.

The model developed so far consists of the diagrams displayed
in Figs.~1 and 2 (using the complex propagators discussed above). 
These diagrams make up the ``effective''
$NN$ potential, Eq.~(\ref{eq_eff}), that is applied in
the scattering equation, Eq.~(\ref{eq_t}), to determine
the $NN$ $T$-matrix, from which phase parameters and
observables can be calculated.
It is wellknown that models of this 
kind~\cite{FT84,HM87,Els88b,foot1} are able to
describe $NN$ scattering up to about 1 GeV in semi-quantitative
terms. Using the parameters listed in Table~I,
phase shifts and inelasticity parameters are predicted as shown in Fig.~3
and mixing parameters as in Fig.~4.
It is seen that several phase shifts are predicted quantitatively,
notably the $S$ waves; others are semi-quantitative, like
the $P$ waves which, typically, show too much attraction
at intermediate energies. The cusps that are known to be the
signature of the $\Delta$ threshold~\cite{Ver82} also show up clearly: 
the shape of the $^1D_2$ phase shift is well reproduced
while, in $^3F_3$ and $^3P_2$, only the trends are right.
Inelasticities are by and large described well, but in the
crucial cases, namely, $^1D_2$ and $^3F_3$ 
the inelasticity is predicted too small. 
This is a wellknown problem~\cite{FT84,HM87,Els88b}.
However, over-all we perceive the agreement between predicted and empirical
phase parameters displayed in Figs.~3 and 4 as sufficient to conclude
that we have a satisfactory understanding of $NN$ scattering
up to about 1 GeV in terms of a relativistic meson model
extended by the $\Delta(1232)$ resonance.

\section{The energy regime above 1 GeV}

With this section, we will start to investigate the issue
to which extend the relativistic meson model
of the previous section can be stretched beyond 1 GeV.
We will consider, first, the most inclusive observables,
namely, total cross sections. The predictions
for the total elastic and the total (i.~e., elastic plus inelastic)
cross sections are shown in Fig.~5 for energies up to 5 GeV
laboratory energy~\cite{foot2}.
From the figure, one can make two important observations:
\begin{itemize}
\item
The predicted inelasticity (difference between the dashed and
solid curve in Fig.~5) is substantially too small above 1 GeV.
\item
The predicted elastic cross section rises with energy
while, empirically, it drops.
\end{itemize}
The lack of inelasticity is not unexpected since, for
$T_{\rm lab} > 1$ GeV, the effectiveness of the
$\Delta(1232)$ resonance is diminishing, while
other inelastic processes enter the picture which are not included in our
model.
However, the number of inelastic channels that open above
1 GeV increases so rapidly with energy~\cite{PDG02}
that it would be inefficient to take
them into account one by one.
Except for the shoulder
around $T_{\rm lab} \approx 800$ MeV which is due
to the $\Delta$ resonance, the inelastic
cross section is smooth and does
not show any structures that would be indicative for
the outstanding role of another meson-nucleon
resonance or a particular inelastic channel.
Hence, a picture of many overlapping resonances and
inelastic channels emerges which
suggests that
further inelasticity can be pragmatically described
by a smooth optical potential.

In configuration space ($r$-space) calculations,
the following form has been used
for a nonrelativistic optical potential~\cite{Ger98}:
\begin{equation}
\tilde{V}_{\rm opt}(r,s) = \left[\tilde{V}_0(s) + i\tilde{W}_0(s)\right] 
\exp\left(-\frac{r^2}{a^2}\right)
\, ,
\label{eq_optr}
\end{equation}
where, as before, $s$ denotes the square of the total c.m.\
energy.
The Gaussian shape is suggested by the geometrical picture proposed
by Chou and Yang~\cite{CY68}, in which two colliding nucleons
are described as extended objects made from some kind of
hadronic matter that has a distribution similar to the charge
distribution. The proton electromagnetic form factor is well represented
by a Gaussian. 

Since we work in momentum space, we Fourier transform
Eq.~(\ref{eq_optr}), yielding
\begin{equation}
\hat{V}_{\rm opt} (k,s) 
= \left( \frac{\sqrt{\pi} a}{2\pi} \right)^3 
\left[ \tilde{V}_0(s) + i\tilde{W}_0(s) \right] 
\exp\left(-\frac{k^2 a^2}{4}\right) \,.
\label{eq_optk}
\end{equation}
This is a nonrelativistic scalar. However, our approach is relativistic
and, therefore, all contributions must have a proper
Lorentz-Dirac structure. In analogy to the nonrelativistic approach,
the obvious choice is a Lorentz scalar~\cite{HR90} which
we define as follows (using the formalism of Refs.~\cite{Mac89}
and \cite{Mac93}):
\begin{equation}
\langle \vec{q'} \lambda_{1}'\lambda_{2}'
|\bar{V}_{\rm opt}|
{\vec q}\lambda_{1}\lambda_{2}\rangle
 = \hat{V}_{\rm opt} (k,s) 
\left[ \bar{u}(\vec{q'},\lambda_{1}')    u({\vec q},\lambda_{1}) \right]
\left[ \bar{u}(-\vec{q'},\lambda_{2}')   u(-{\vec q},\lambda_{2}) \right]
\,,
\label{eq_optrel}
\end{equation}
where 
$\lambda_1, \lambda_2$ 
($\lambda_1', \lambda_2'$) 
denote the helicities of the two ingoing (outgoing)
nucleons and 
$\vec{q}$
($\vec{q'}$) 
are the corresponding relative momenta in the c.m.\ system;
$k \equiv | \vec{q'} - \vec{q}|$
is the magnitude of the momentum transfer between the interacting
nucleons.
The Dirac spinors in helicity representation are given by
\begin{eqnarray}
u({\vec q},\lambda_1)&=&\sqrt{\frac{E_q+M}{2M}}
\left( \begin{array}{c}
       1\\
       \frac{2\lambda_1 |{\vec q}|}{E_q+M}
       \end{array} \right)
|\lambda_1\rangle
\: ,
\\
u(-{\vec q},\lambda_2)&=&\sqrt{\frac{E_q+M}{2M}}
\left( \begin{array}{c}
       1\\
       \frac{2\lambda_2 |{\vec q}|}{E_q+M}
       \end{array} \right)
|\lambda_2\rangle
\: ,
\end{eqnarray}
which are normalized such that
\begin{equation}
\bar{u}({\vec q},\lambda) u({\vec q},\lambda)=1
\, ,
\end{equation}
with $\bar{u}=u^{\dagger}\gamma^{0}$.

Instead of fitting the parameters of the optical potential,
$\tilde{V}_0(s)$ and $\tilde{W}_0(s)$,
separately for various single energies~\cite{Ger98,HR90}, we find
it physically more reasonable to choose a smooth, analytic
function of $s$ with the correct high-energy behavior built in:
\begin{equation}
\tilde{V}_0(s) + i\tilde{W}_0(s) = \left\{
       \begin{array}{ll}
       0         & \mbox{for $T_{\rm lab} \leq T_{\rm lab}^{(0)}$} \\
       \left( \frac{s-s_0}{4M^2} \right) [V_0+iW_0]
                     & \mbox{for $T_{\rm lab} > T_{\rm lab}^{(0)}$} \,,
       \end{array}
   \right.
\label{eq_opte}
\end{equation}
where $s_0=2M(2M+T_{\rm lab}^{(0)})$ and $T_{\rm lab}^{(0)} = 0.8$ GeV.
This parametrization implies that the optical potential is proportional
to $s$ for large energies ($s \gg s_0$) which leads to constant
total cross section predictions in the energy range 10 to 100 GeV,
consistent with experiment~\cite{PDG02}.

In summary, to generate the additional inelasticity required for
the total cross sections above about 1 GeV, we add to the model
developed in Sec.~II the relativistic optical potential
defined by Eqs.~(\ref{eq_optrel}),
(\ref{eq_optk}) and (\ref{eq_opte}).

We now turn to the other deficiency that we are observing
in Fig.~5, namely, the rise of the elastic cross section with
energy, which contradicts experiment.
One-particle exchange creates amplitudes that have the basic
mathematical structure
\begin{equation}
\bar{V}_\alpha \propto \frac{s^J}{t-m_\alpha^2}
\end{equation}
where $J$ denotes the spin of the exchanged particle
and $t$ is the square of its four-momentum.
The vector mesons $\rho$ and $\omega$ have $J=1$
and, therefore, create total cross sections that rise
with $s$. This is the basic reason for the rising
cross sections seen in Fig.~5.

This failure of the one-particle exchange picture at high energies
has been known since the late 1950's, when data of sufficient energy
became available to reveal this problem.
In an attempt to solve this problem,
Regge theory~\cite{AR65,Per74,Col77}
was developed in the early 1960's
which, indeed, is able to reproduce the general energy behavior
of two-body cross sections, correctly.
In the Regge model, one-particle exchange is replaced
by the exchange of a Regge pole,
which is an infinite series of particles with the same
spin, isospin, and strangeness, aligned along
a Regge trajectory. Regge trajectories are named by
their first, best-known family member:
there exists a $\rho$ and an $\omega$ trajectory.

Based upon these historical developments,
it might be appealing to replace the one-rho
and one-omega exchanges in our model by the
corresponding Regge trajectories.
However, there are reasons why we should not resort
to such drastic measures.
From a modern point of view, Regge theory is essentially
a phenomenology for very high energies. It is most appropriate
above about 10 GeV which is beyond the energies that we are intersted in.
This fact reveals the greatest dilemma of the energy
regime between 1 and 10 GeV: there exist well-tested models
below 1 GeV (meson model) and above 10 GeV (Regge model), 
however in-between, the established models are partially inadequate
and no alternatives have been proposed.

Another problem with Regge theory is that
it does not make any predictions for the spin-dependence
of the interaction, which is one focus of this study (see below).

For the reasons discussed, we will not switch to Regge theory. 
Instead, we will modify our model in the spirit of Regge theory.
Of the Regge trajectories, we will only keep the first family
member. The one-particle exchange amplitude of this first member 
will be modified such
that the main effect of the rest of the trajectory is taken
into account. As discussed, this main effect 
is that it removes the
wrong energy behavior from the amplitude.
Thus, we apply to the one-omega and one-rho exchange
amplitudes a factor that divides
the wrong energy dependence out:
\begin{equation}
\bar{V}_\alpha  \longmapsto  \frac{s_0}{s} \bar{V}_\alpha
\,,
\label{eq_regge}
\end{equation}
for $\alpha=\rho,\omega$ and $s > s_0$ with $s_0$ as defined
below Eq.~(\ref{eq_opte}). For $s \leq s_0$, there are no
changes. The modification, Eq.~(\ref{eq_regge}),
is applied to the $\rho$ and $\omega$ exchanges of
$V_{NN,NN}$ and the $\rho$ exchanges of
$V_{NN,N\Delta}$ and $V_{NN,\Delta\Delta}$
[cf.\ Eq.~(\ref{eq_eff})].

Including the optical potential, Eq.~(\ref{eq_optrel}),
and the modification of the vector meson amplitudes,
Eq.~(\ref{eq_regge}), we obtain the total cross section
predictions displayed in Fig.~6.
The elastic cross section now shows the correct energy
behavior and the inelasticity (and the total cross
section) is of the right size and energy dependence. Thus, based upon
a few physically reasonable assumptions, it is fairly
easy and straightforward to describe the $pp$
total cross sections above 1 GeV.

\section{Spin observables}
In this section, we turn to $pp$ spin observables. 
We will compare predictions by the model developed in
the previous two sections (and variations thereof) to data 
for five representative energies 
in the range 400 MeV to 2500 MeV.
Besides differential cross sections, $d\sigma/d\Omega$, and analyzing
powers, $A_N$, we will consider the spin correlation coefficients
$A_{NN}$, $A_{SS}$, $A_{SL}$, and $A_{LL}$, for which (except
for $A_{LL}$) precise
data have been taken by the EDDA group 
~\cite{EDDA1,EDDA2,EDDA3}, and for $A_{NN}$, $A_{SL}$ at SATURNE~\cite{All98,All01}.
Since the differences between the two experimental data sets are small
as compared to the difference between theory and experiment,
we will subsequently compare only to the EDDA data.

The predictions by the relativistic meson model
presented in Sec.~II, as modified in Sec.~III,
are shown by the solid curve in Fig.~7. The dashed curve
in the figure is obtained when the optical potential,
Eq.~(\ref{eq_optrel}), and the corrections to
vector meson, Eq.~(\ref{eq_regge}), are left out.
Finally, the dotted curve is based upon the GWU (formerly VPI)
phase shift analysis SP03~\cite{SP03}.
Since a phase shift analysis is just an alternative way
of representing data, the dotted curve follows, in general,
the data included in Fig.~7.
The exception is $A_{LL}$, where no data exist
and where, therefore, the analysis represents the only
empirical information to compare with.

Since we expect the meson model to be right at least for
low energies, it is comforting to see that
at 400 MeV there is generally
good agreement between theory and experiment
for all observables shown.
Consistent with the predictions for total cross
sections discussed in Sec.~III, the differential cross sections 
come out too large above 1 GeV when the modifications
introduced in Sec.~III are not applied (dashed curve). 
Including those modifications (solid curve)
yields a better agreement
for all energies up to 2.5 GeV, for differential cross sections.

However, for spin observables, the agreement is much
less satisfactory.
Already at 800 MeV,
the analyzing power is predicted substantially too high,
which is probably associated with the fact that the $^3P_2$
phase shift is predicted too large above 650 MeV
(cf.\ Fig.~3); 
note that only spin-triplet partial waves enter the 
amplitudes describing $A_N$.

At higher energies, the corrections necessary to improve
the cross sections enhance $A_N$ contrary to the data.
So, overall, the analyzing power is predicted persistently
too large.

In the case of the spin correlation parameters,
the correction applied to vector-meson exchange,
Eq.~(\ref{eq_regge}), and the optical potential
provide effects that point in the right direction.
Nevertheless, the best one can say is that
theory and experiment agree in the trends that
the spin correlation coefficients show as a function of
angle. But, in quantitative terms, there are large discrepancies.

The parameters of our model are the meson-baryon coupling constants
and the cutoff masses, which parametrize the meson-baryon form
factors. With the exception of the $\pi NN$ and $\pi N \Delta$
coupling constants, these parameters are only losely constrained
by information from other sources. Consequently, the parameter set
that we have used so far 
is not the only choice that can be made.
Therefore, we have varied all parameters within the ranges
given in Table~I.
These ranges represent educated estimates of the uncertainties.
The result of this very comprehensive investigation of a systematic 
variation of all parameters can be summarized as follows:
It is not possible to obtain a fit
of all observables at all energies considered
that is substantially better than the one shown in Fig.~7.

To obtain further insight into the nature of the problem,
we have investigated the question,
if it is at least possible to fit single observables at
single energies. We selected a few representative cases
and found for all of them that it was, indeed, possible
to find a
combination of parameters that resulted in a good fit
of the single observable chosen.
To illustrate this point, we show in Fig.~8 the case where the
prediction for $A_N$ at 1.8 GeV 
is substantially improved. However, it is clearly
seen that the fit of all the other observables is now, in general, worse
than in Fig.~7, including $A_N$ at lower energies.
Thus, the quantitative fit of just one observable at one energy and for
a limited range of angles
cannot be perceived as a confirmation of the validity
of the meson model at high energies~\cite{HM84}. On the 
other hand, the fact that all observables can be fitted  separately by
some individually adjusted combination of parameters implies
that our model does contain all the types of spin-dependent forces necessary
to describe the $NN$ amplitudes.
What fails is the energy- and angle-dependence.
While for low energies (below $\approx 0.8$ GeV)
the meson model generates the correct energy-dependence
for the strength of the various spin-dependent components,
this energy-dependence becomes increasingly wrong
when proceeding to higher energies.

\section{Conclusions}

In this paper, we have studied $NN$ scattering above 1 GeV laboratory energy.
We started from a model that is based upon relativistic meson-exchange,
complemented by the $\Delta(1232)$ isobar, and reproduces $NN$ scattering
up to about 1 GeV satisfactorily. We have then extrapolated this model above
1 GeV. At those higher energies, characteristic deficiencies in the total
and differential cross sections show up that are easy to fix.
The lack of inelasticity is mended by introducing an optical
potential of the shape of the proton form factor. The well-known 
wrong high-energy behavior
of vector-meson exchange becomes noticable already around 1.2 GeV.
In the spirit of Regge theory, we apply a factor $s_0/s$ to vector
mesons with the consequence that the elastic cross sections and their
energy behavior are predicted correctly.

An important focus of our study have been spin-observables of
$pp$ scattering. Due to recent experiments conducted by the
EDDA group~\cite{EDDA2,EDDA3}, data on spin correlation coefficients
(besides analyzing powers)
up to 2.5 GeV are now available 
for a broad range of angles.
Comparison of our predictions with these data confirms the
well-known fact that a correct
reproduction of cross sections by no means implies a correct
description of spin observables. Even the ``simplest'' spin observable,
namely, the analyzing power $A_N$, poses a challenge to theory
which predicts $A_N$ persistently too large.
Concerning the more sophisticated spin correlation parameters,
the only encouraging statement that can be made is that the characteristic
trends of these observables as a function of angle
come out about right. But there is no quantitative agreement.
Varying the parameters of the model (coupling constants and
cutoff parameters) over a wide range does not improve
the over-all quality
of the description of the data.

In conclusion, we do not have a quantitative understanding
of the spin-dependence of the $NN$ interaction above 1 GeV.
The meson model, which is so successful at low energies, 
becomes increasingly inadequate above 1 GeV. This fact is revealed
most clearly by spin observables.

It is tempting (since plausible) to interpret the gradual failure of the meson model
with increasing energy as an indication that pQCD
is becoming the valid approach at higher energy. 
Unfortunately, this suggestion is not correct.
The implications of pQCD for $NN$ elastic scattering
have been worked out carefully in Ref.~\cite{QCD79} and
the prediction clearly is:
\begin{equation}
A_{N} = 0
\label{eq_ann0}
\end{equation}
at all angles.
This is not what we see in the data.
In fact, $A_N$ was measured up to laboratory energies
of 28 GeV by Alan Krisch and the Michigan group~\cite{FK81}
and there are no indications for a decline of $A_N$
even at those large energies.

Assuming massless, effectively free quarks, the helicities
of the quarks are conserved, which implies for
the spin-correlations parameters~\cite{QCD79}:
\begin{eqnarray}
A_{NN} & = & - A_{SS} \,,
\label{eq_ann1a}
\\
A_{SL} & = & 0 \,,
\label{eq_ann1b}
\end{eqnarray}
for all angles. Applying the quark-interchange model
to this scenario, yields the specific predictions~\cite{QCD79}
\begin{eqnarray}
A_{NN} (90^0) & = & \frac13 \,,
\label{eq_ann2a}
\\
A_{LL} (90^0) & = & A_{SS} (90^0) = - \frac13 \,,
\label{eq_ann2b}
\end{eqnarray}
which (as it should) satisfies the model-independent sum
rule:
\begin{equation}
A_{NN} (90^0) - A_{LL} (90^0) -  A_{SS} (90^0) = 1 \,,
\end{equation}
where the angle is measured in the c.m.\ system.

Accidentally, the data at 800 MeV and above displayed in
Fig.~7 agree roughly with Eqs.~(\ref{eq_ann1a})
and (\ref{eq_ann1b}). However, we should not interpret
this as a signature of pQCD. The Michigan group~\cite{FK81}
measured $A_{NN}$ at $90^0$ up to 12 GeV
and found strong variations with energy, and a value
of about 0.6 at 12 GeV which disagres
by a factor two with Eq.~(\ref{eq_ann2a}).
Moreover, the best-founded implication of pQCD is a
vanishing analyzing power and, therefore, if this condition
is not met, we are not in pQCD territory.

In lack of a calculable high-energy theory,
one may consider to resort to traditional high-energy 
phenomenology. The Regge model complemented by
Pomeron exchange is the most successfull
phenomenology for the description of hadron-hadron cross sections
above 10 GeV laboratory energy~\cite{DL92,Mat94}. 
However, the main problem that
we are facing in this study are spin observables.
To our knowledge, the exact implications of Regge theory for the
spin-dependence of the $NN$ interaction has never been
worked out, since most of the work on Regge theory was done in the 1960s
when polarization data at high energy were not available.
The work by Rijken~\cite{Rij85}
on low-energy implications of Regge theory suggests
that Regge theory predicts a spin-dependence similar
to the OBE model.
If true, then our model contains already all the
spin-dependence that a Regge theory would produce.
On the other hand, one may also raise objections concerning
the use of Regge theory:
In general, the
Regge model is perceived as appropriate in the
energy regime above 10 GeV and, so, it is questionable if
it is the right phenomenology for energies at a few GeV
which is our focus.

In summary, the energy region between 1 and 10 GeV poses a
serious problem:
the energies are too high for typical nuclear physics
approaches (like, chiral perturbation theory or meson models) 
and too low for typical high-energy theories. 
In this sense, the region 1-10 GeV is the true
``intermediate energy'' region.
The transition character of this region may be the crucial
underlying reason why, so far, any attempt to explain the data
has just opened Pandora's Box.

In the late 1970s and early 1980s, when
the first measurements by Alan Krisch and co-workers~\cite{FK81}
of unexpectedly
large analyzing powers and transvers spin-correlation
coefficients in $pp$ scattering at high energies and
large angles had become known to the community, 
a flurry of theoretical activity evolved~\cite{QCD79,HM84,BT88,HR90,foot3}. 
However, none of the many theoretical papers really
solved the problem of the spin-dependence of the $NN$ interaction
at higher energies and, after a while, the community 
simply lost interest in the subject.
With the new data on spin-correlation coefficients~\cite{EDDA3}
the problem is more apparent than ever.
The fact that we do not have a precise understanding
of the $NN$ interaction above 1 GeV is a serious
problem that deserves the attention of
the community. We need new ideas and much more theoretical work.

\acknowledgements
This work was supported by the U.S. National Science
Foundation, Grant No.~PHY-0099444, the BMBF (Germany), Grant No. 06HH152, and 
the FZ J\"ulich (Germany), FFE 41520732.

\newpage

\begin{table}
\caption{Meson parameters. ($J$, $P$, and $I$ denote spin, parity, and isospin
of mesons.)}
\begin{tabular}{cccccc}
Meson  & $J^P$ & $I$ & $m_\alpha$ (MeV)$^a$ &  $g^{2}/4\pi$ or $f^{2}/4\pi$$^a$
 &  $\Lambda_{\alpha}$ (GeV)$^a$ \\
\hline 
\multicolumn{6}{c}{$NN\alpha$ vertices}\\
$\pi$ & $0^-$ & 1 & 138.03  &  14.4$^b$ & 1.6 (1.1 - 2.1)  \\
$\eta$ & $0^-$ & 0 & 548.8  &   2.0 (1.0 - 3.0)$^b$ & 1.5 (1.3 - 1.7)  \\
$\rho$ & $1^-$ & 1 & 769.0  &  1.1 (0.3 - 1.1)$^c$ & 1.3 (1.0 - 1.8)  \\
$\omega$ & $1^-$ & 0 & 782.6 & 23.0 (17.0 - 31.0)$^d$ &  1.5 (1.3 - 2.1)  \\
$\sigma^e$ & $0^+$ & 0 & 500.0 (300.0 - 800.0)&   3.676 (1.5 - 5.5) &  1.5 (1.1 - 1.9) \\
\hline 
\multicolumn{6}{c}{$N\Delta\alpha$ vertices}\\
$\pi$  & $0^-$ & 1 & 138.03 & 0.35 & 0.9 (0.7 - 1.1) \\
$\rho$ & $1^-$ & 1 & 769.0  & 20.45 (16.0 - 28.0)& 1.4 (1.2 - 1.6) \\
\end{tabular}
$^a$Numbers in parentheses state the range of variation.\\
$^b$$g^2_{\alpha NN}/4\pi = (2M/m_\alpha)^2 f^2_{\alpha NN}/4\pi$ is given.\\
$^c$$g^2_\rho/4\pi$ is given; $f_\rho/g_\rho = 6.1$.\\
$^d$$g^2_\omega/4\pi$ is given; $f_\omega/g_\omega = 0$.\\
$^e$The $\sigma$ parameters given in the table
 apply to the $T=1$ $NN$ potential; for the $T=0$
potential, $g^2_\sigma/4\pi=2.5064$ and $m_\sigma=450$ MeV
are used.
\end{table}

\newpage

\begin{figure}
\begin{center}
  \psfig{figure=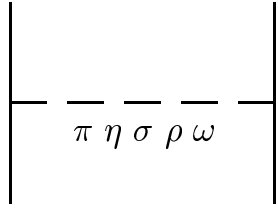,height=4cm}
\end{center}
\caption{One-boson exchange contributions to the $NN$ interaction.
Solid lines represent nucleons and dashed lines are mesons.}
\label{fig_1}
\end{figure}

\begin{figure}
\begin{center}
  \psfig{figure=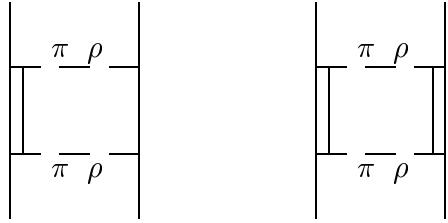,height=4cm}
\end{center}
\caption{Two-meson-exchange box-diagram contributions to the $NN$ interaction 
involving nucleons (solid lines) and $\Delta$ isobars (double lines).
The dashed lines represent $\pi$ and $\rho$ exchange.}
\label{fig_2}
\end{figure}

\pagebreak

\begin{figure}
\begin{center}
  \psfig{figure=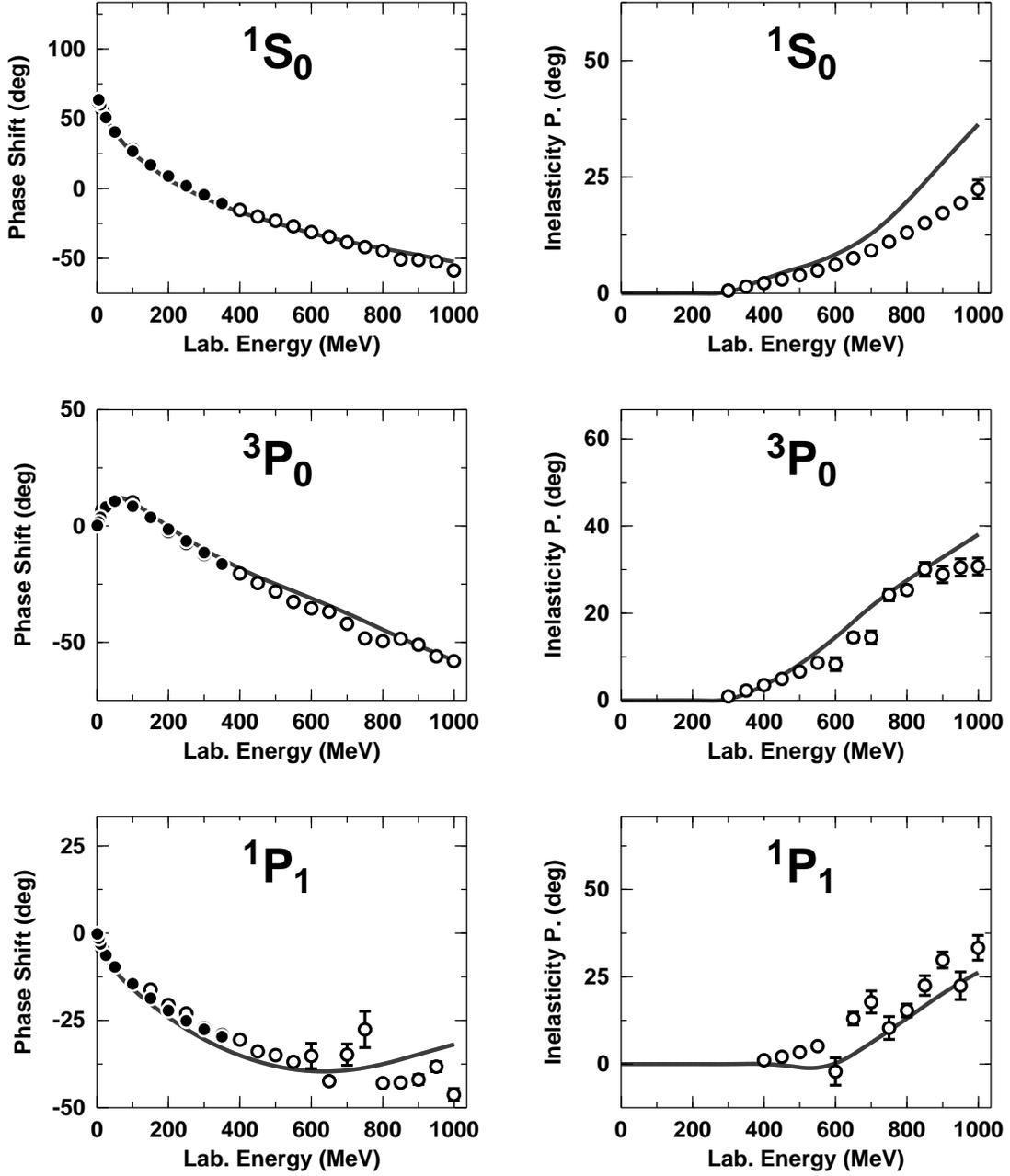,height=18.1cm}
\end{center}
\caption{Phase shifts and inelasticity parameters of $NN$ 
scattering below 1 GeV laboratory energy.
The solid curve represents the predictions by the model described in Sec.~II.
The solid dots show the Nijmegen
multi-energy $np$ phase shift analysis~\protect\cite{Sto93}, and 
the open circles are the GWU (formerly VPI)
single-energy $np$ analysis SP03~ \protect\cite{SP03}.
Arndt-Roper conventions are used for the phase 
parameters~\protect\cite{AR82}.}
\label{fig_3}
\end{figure}

\pagebreak

\begin{center}
  \psfig{figure=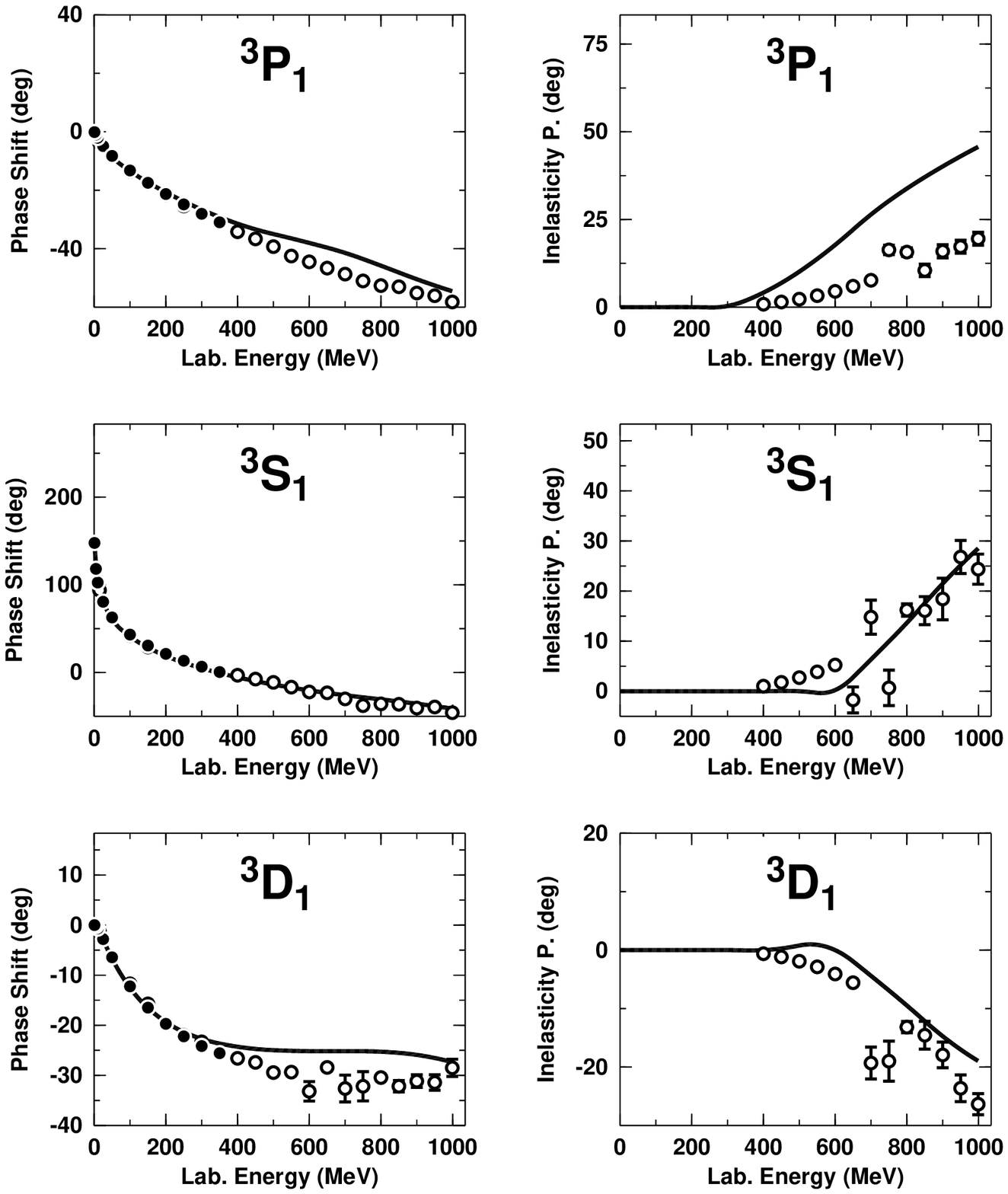,height=18.1cm}
  Fig.~\ref{fig_3} continued.
\end{center}

\pagebreak

\begin{center}
  \psfig{figure=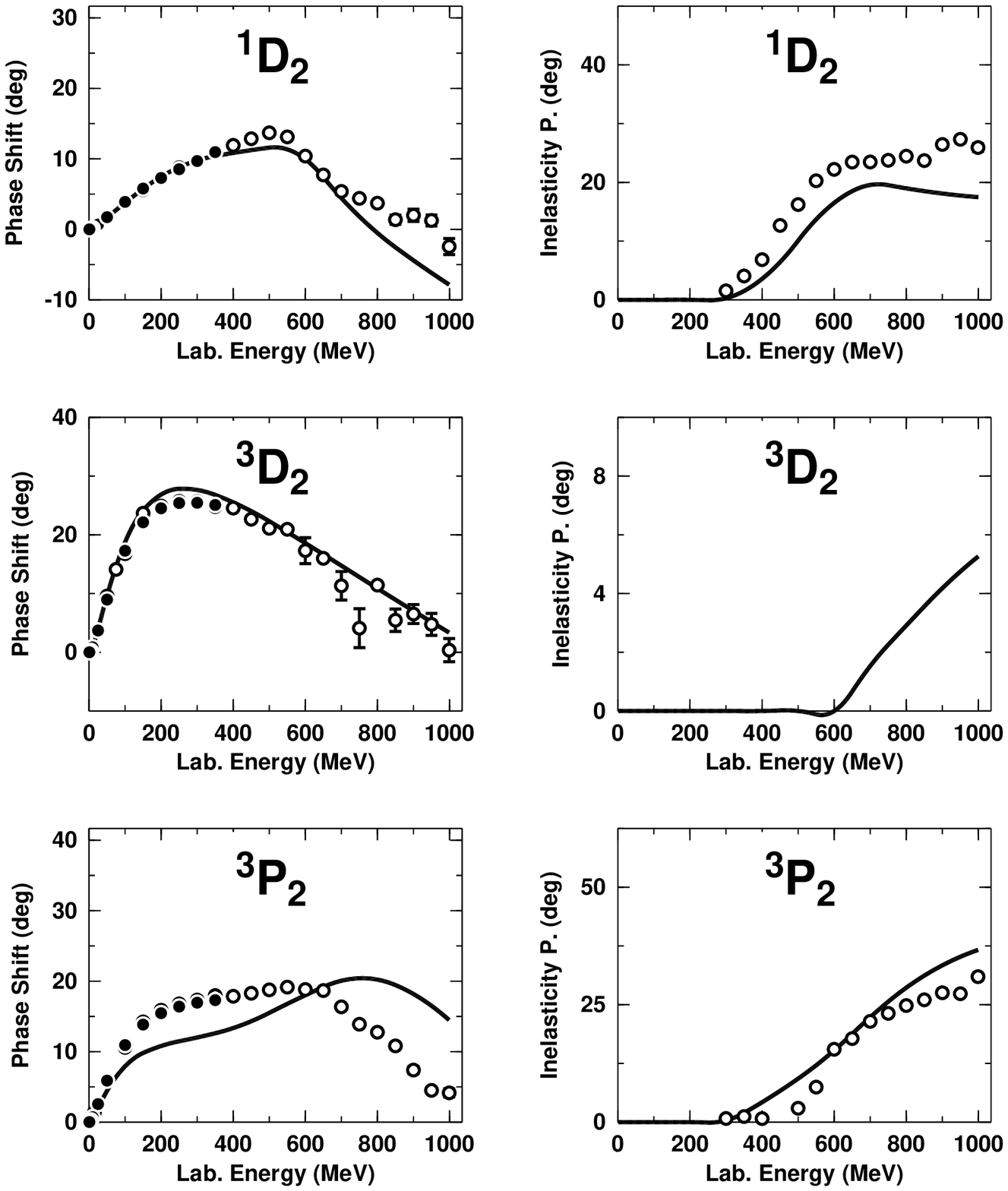,height=18.1cm}
  Fig.~\ref{fig_3} continued.
\end{center}

\pagebreak

\begin{center}
  \psfig{figure=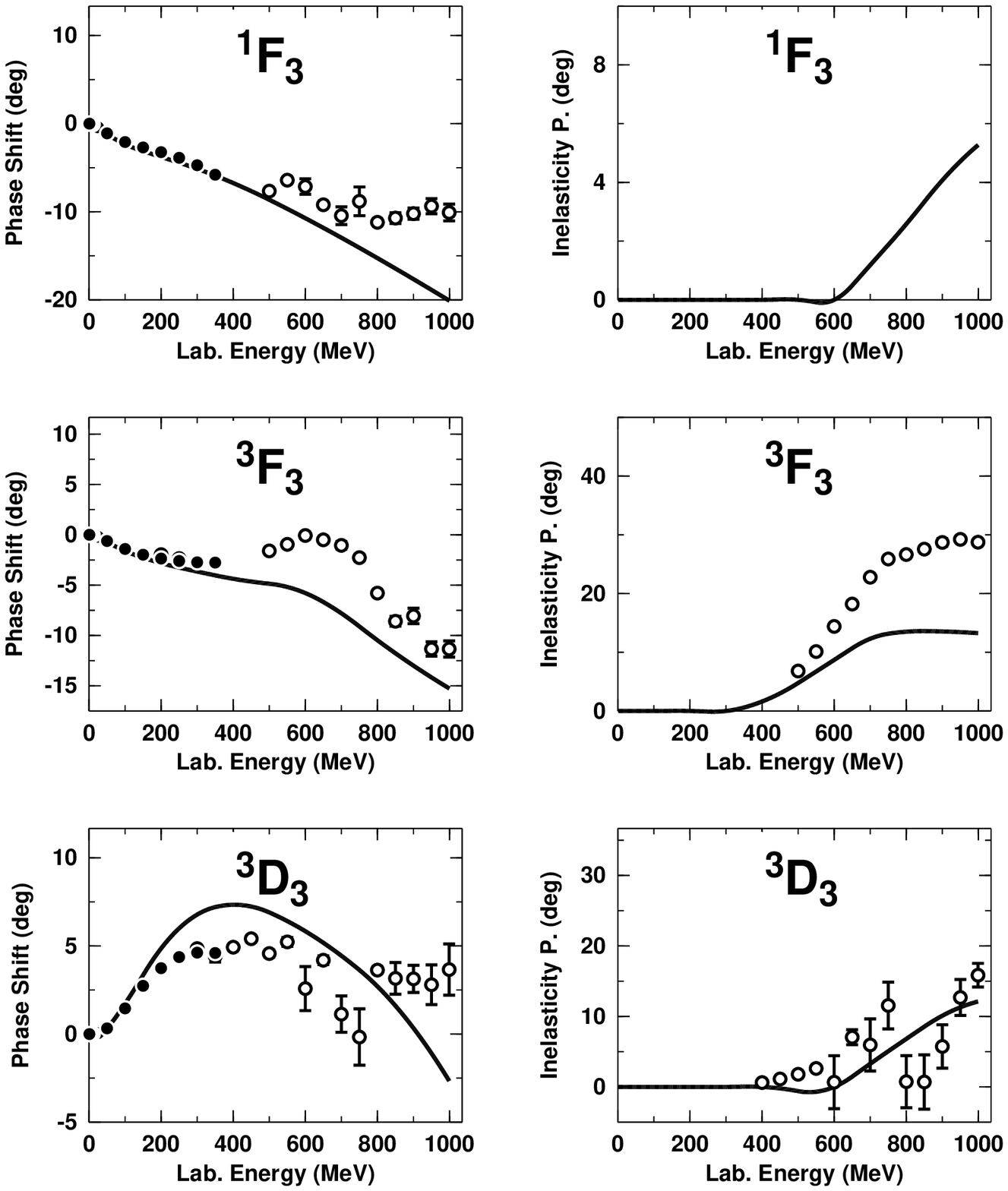,height=18.1cm}
  Fig.~\ref{fig_3} continued.
\end{center}

\pagebreak

\begin{figure}
\begin{center}
  \psfig{figure=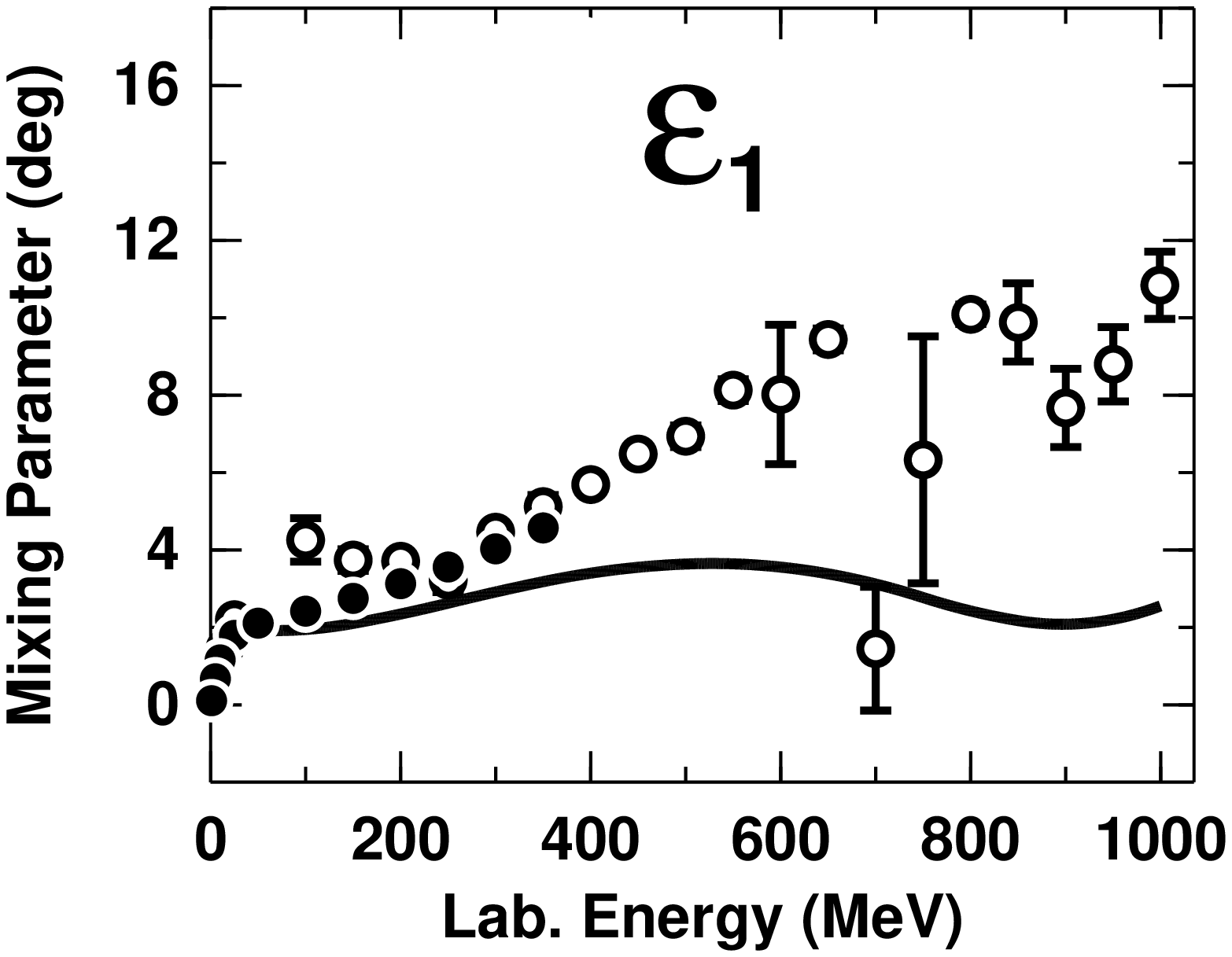,height=5.58cm}
  \psfig{figure=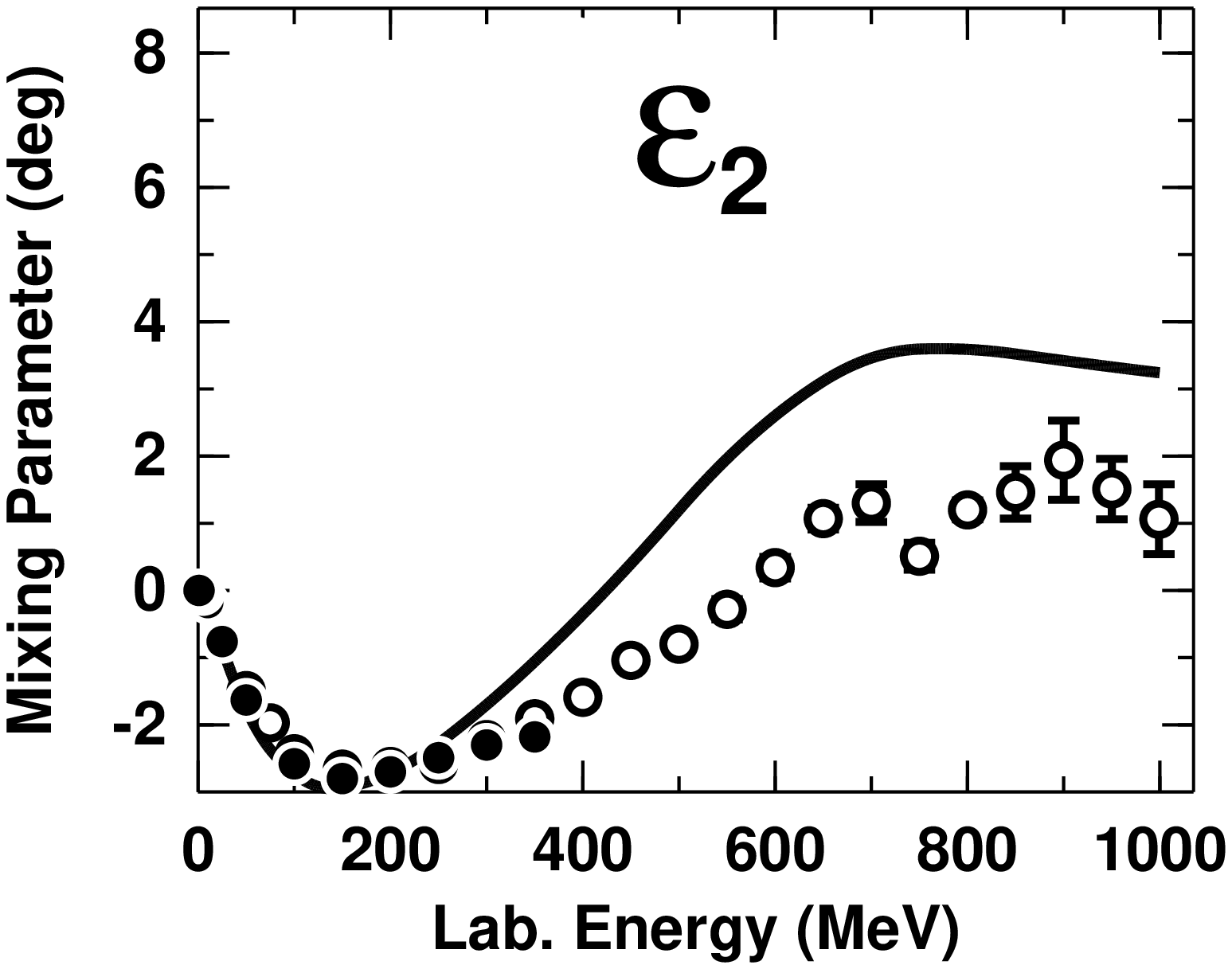,height=5.58cm}

  \psfig{figure=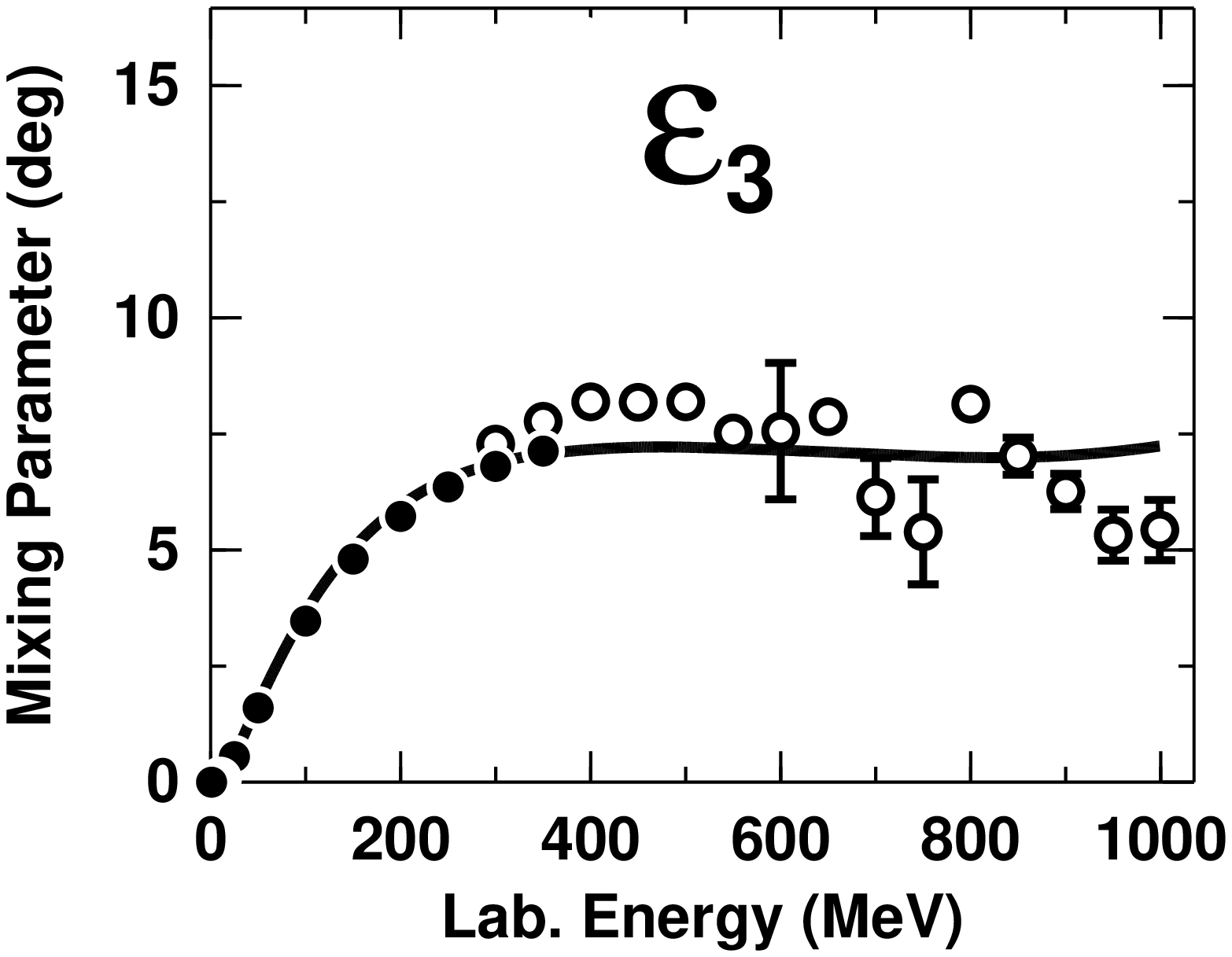,height=5.58cm}
  \psfig{figure=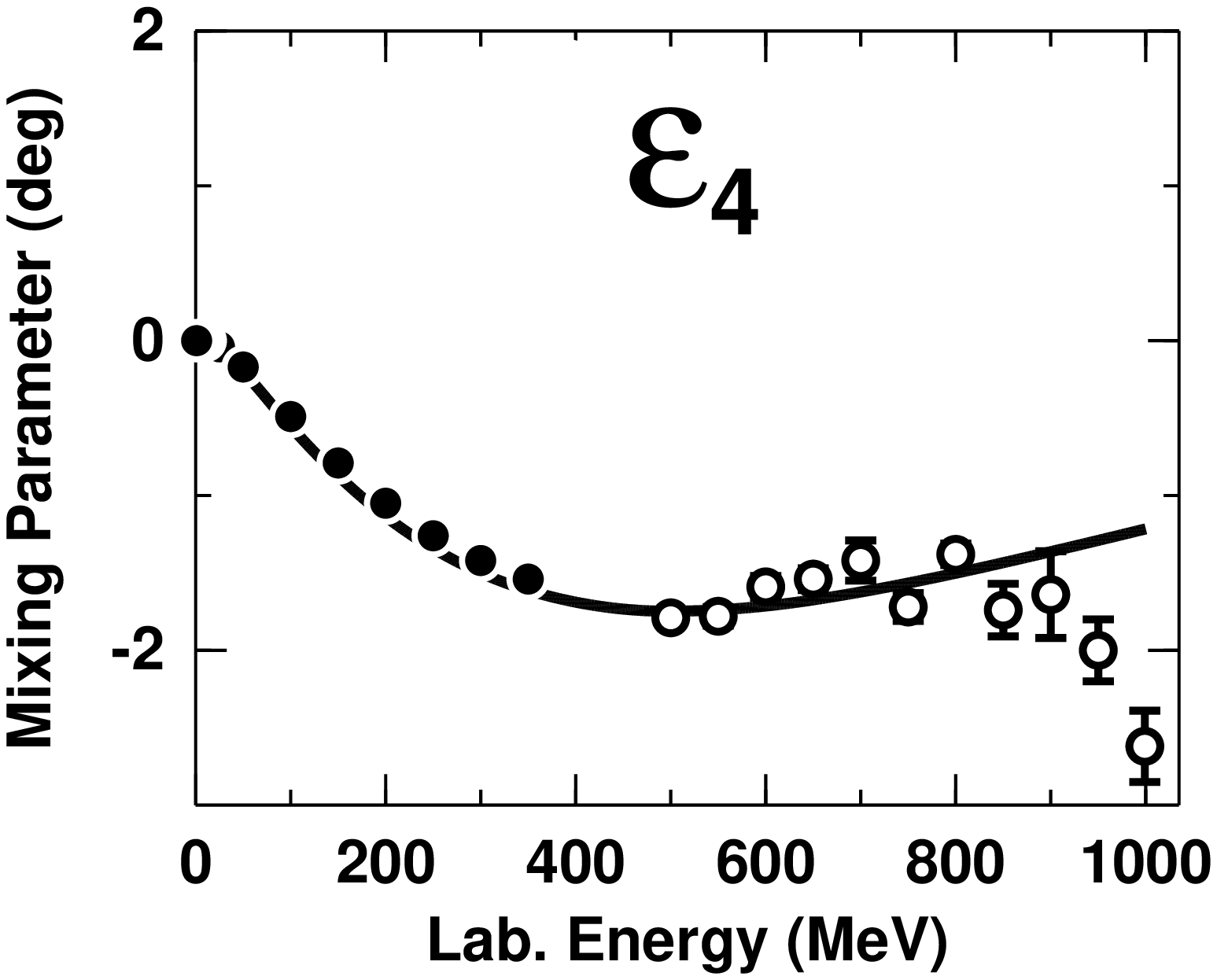,height=5.58cm}
\end{center}
\vspace{.8cm}
\caption{Mixing parameters for $J \leq 4$ and 
laboratory energies below 1 GeV.
The solid curve represents the prediction by the model described in Sec.~II.
The solid dots show the results from the Nijmegen
multi-energy $np$ analysis~\protect\cite{Sto93}, and 
the open circles are the GWU (formerly VPI)
single-energy $np$ analysis SP03~ \protect\cite{SP03}.
}
\label{fig_4}
\end{figure}

\pagebreak

\begin{figure}
\begin{center}
  \epsfig{figure=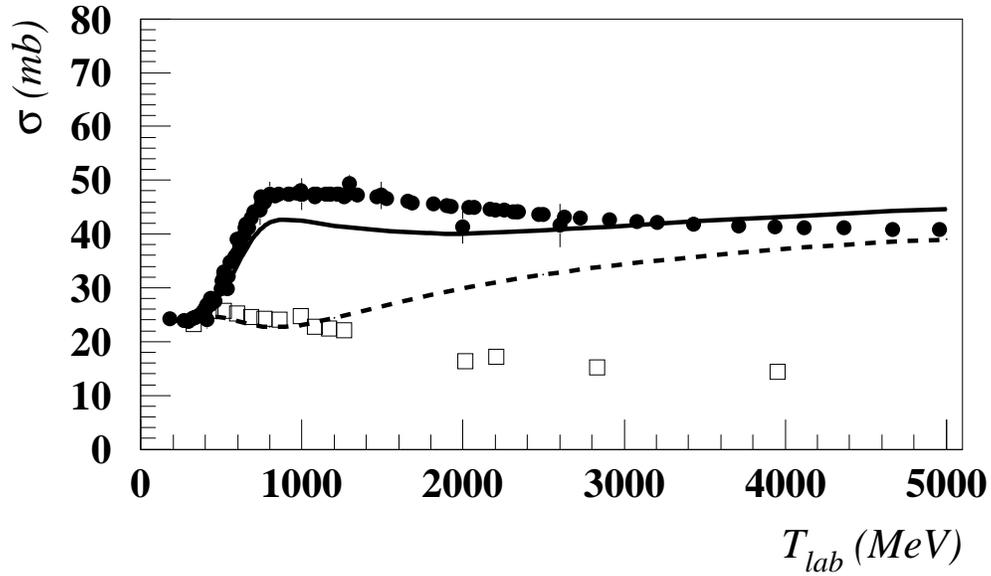,height=8cm}
\end{center}
\caption{Total cross section (solid line) and 
total elastic cross section (dashed line) as predicted
by the relativistic meson model presented in Sec.~II.
The experimental data for total cross sections
are represented by solid symbols, while 
open symbols show the elastic cross section data.
}
\label{fig_5}
\end{figure}

\begin{figure}
\begin{center}
  \epsfig{figure=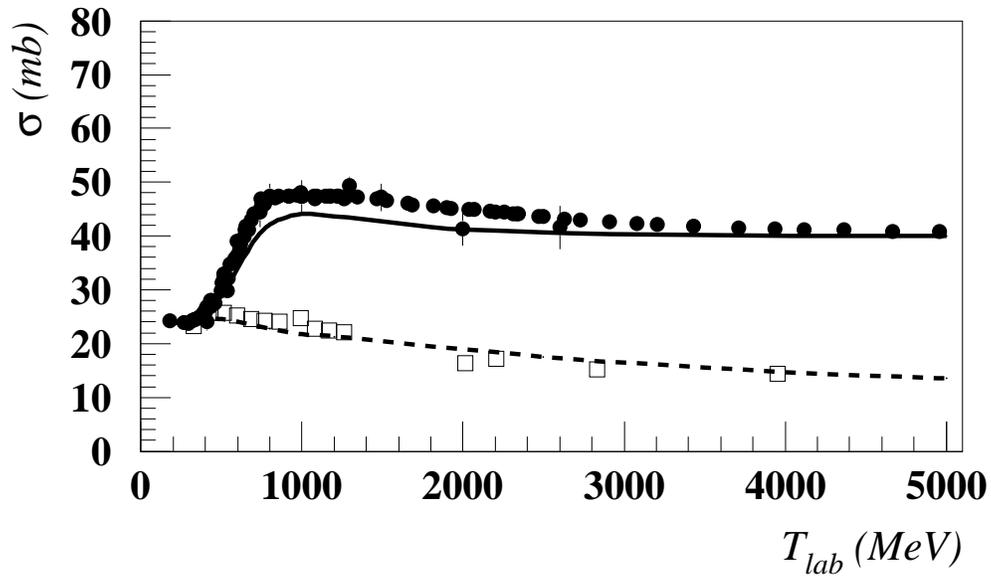,height=8cm}
\end{center}
\caption{Total cross section (solid line) and 
total elastic cross section (dashed line) as predicted
by the relativistic meson model as modified in Sec.~III.
Data as in Fig.~5.
}
\label{fig_6}
\end{figure}

\pagebreak

\begin{figure}
\begin{center}
  \epsfig{figure=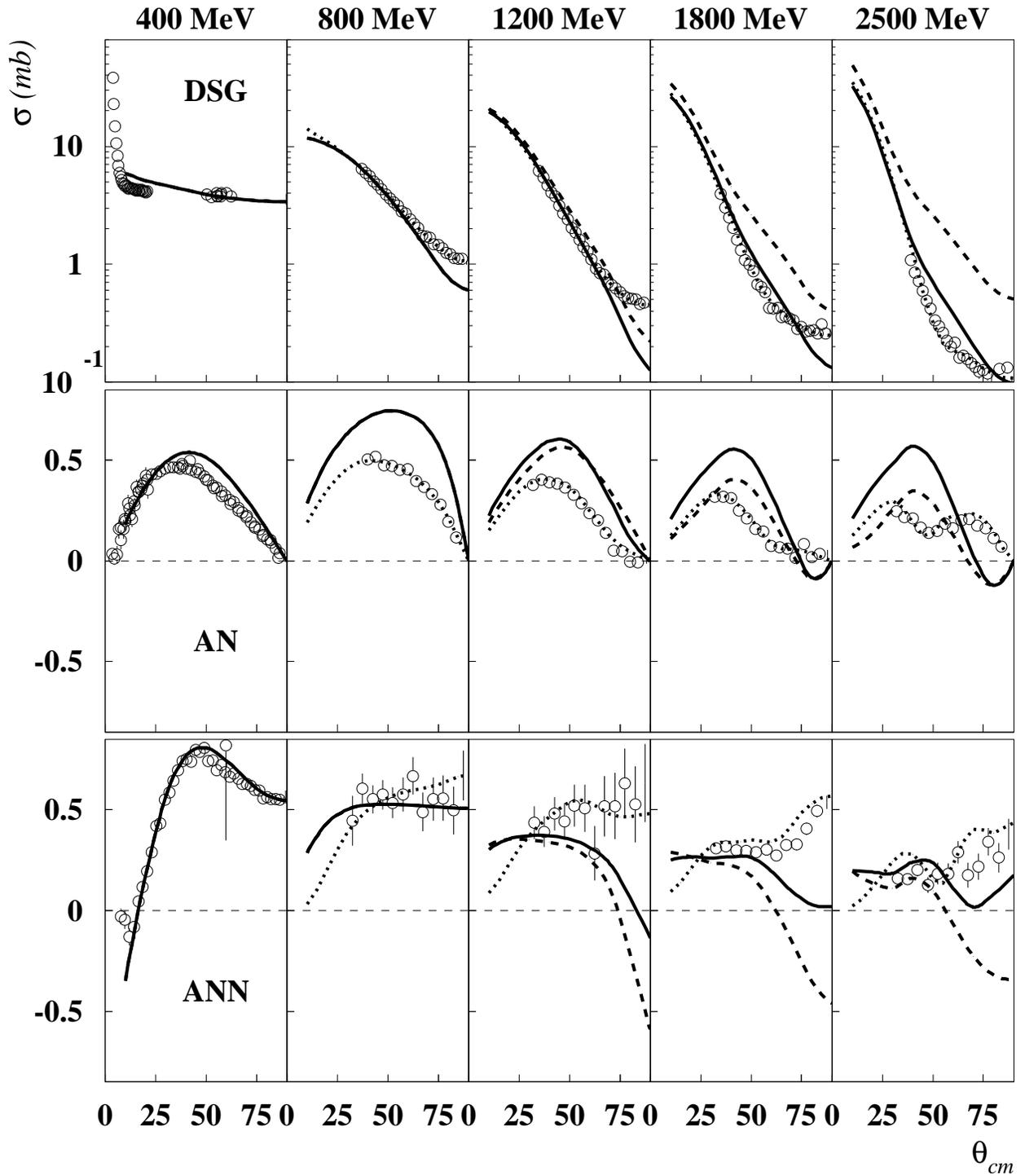,width=17cm}
\end{center}
\caption{Observables of $pp$ scattering as denoted
for five energies between 400 and 2500 MeV lab.\
energies. 
The dashed curve represents the predictions by the model
of Sec.~II, while the solid curve includes
 the modifications of Sec.~III.
The dotted curve is based upon the the GWU (formerly VPI)
phase shift analysis SP03. Data from Refs.~\protect\cite{EDDA1,EDDA2,EDDA3}.
}
\label{fig_7}
\end{figure}

\pagebreak

\begin{center}
  \epsfig{figure=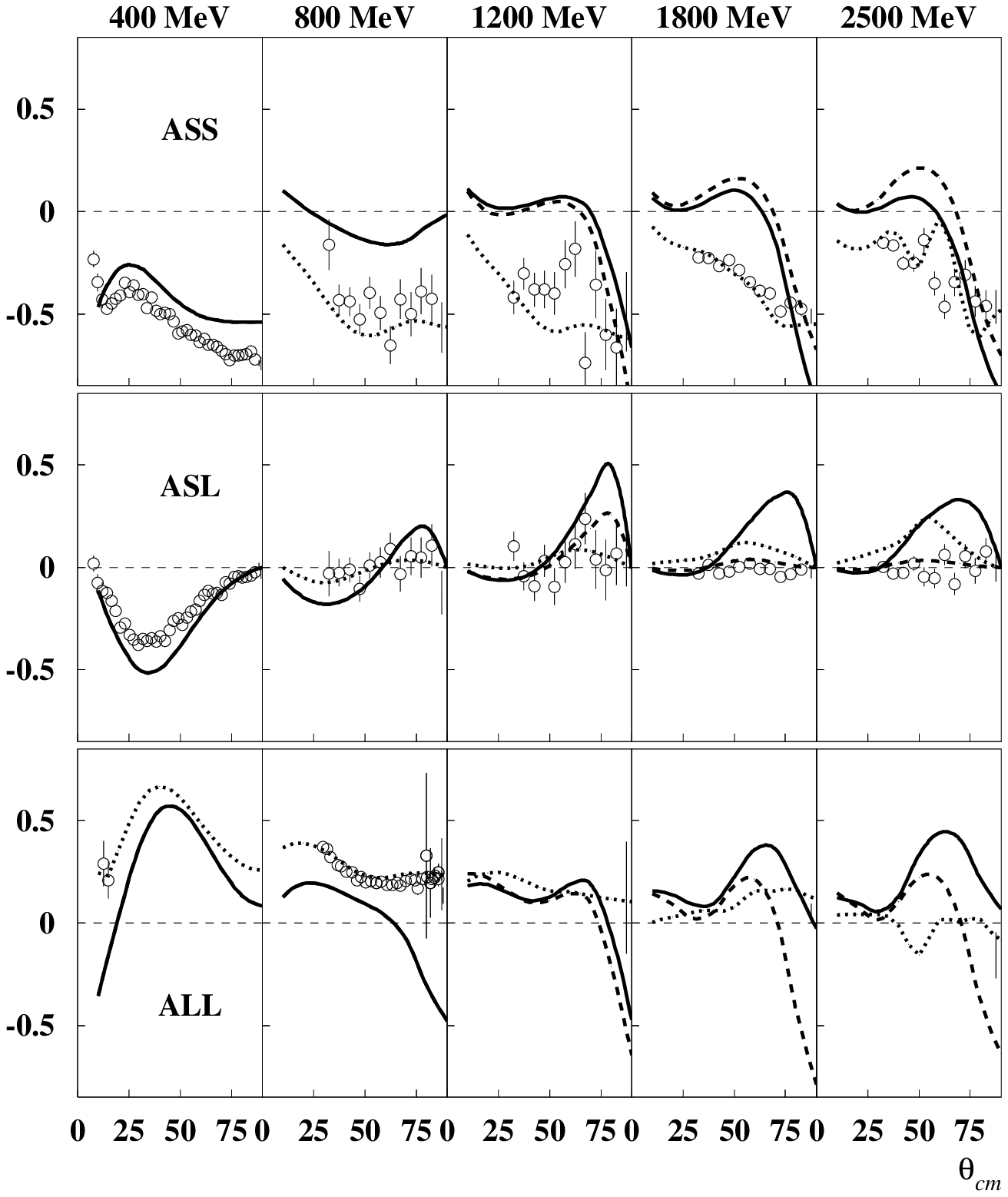,width=17cm}
  Fig.~\ref{fig_7} continued.
\end{center}

\pagebreak

\begin{figure}
\begin{center}
  \epsfig{figure=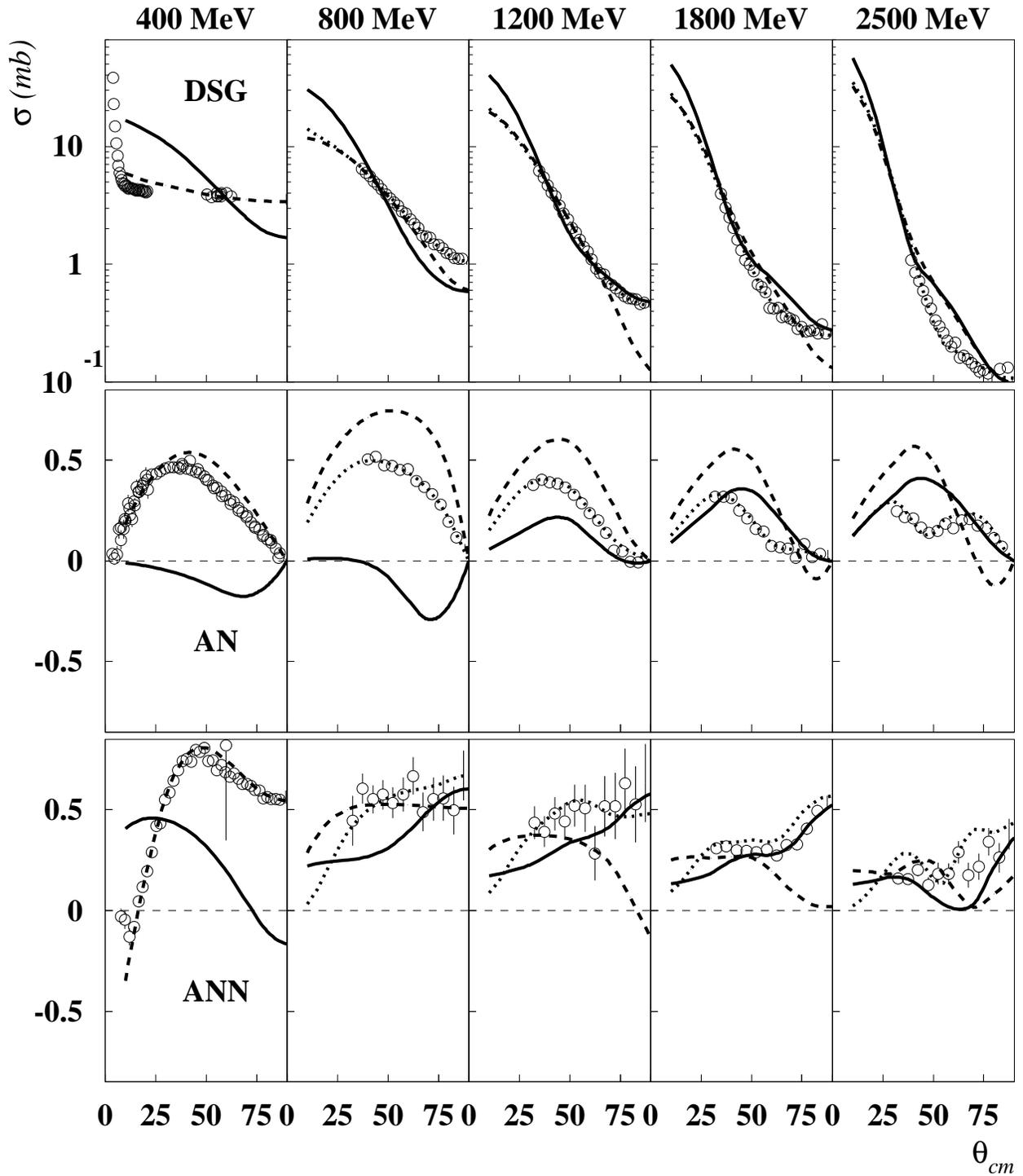,width=17cm}
\end{center}
\caption{The solid curve
represents the prediction by a variation of our model
that yields an improved fit of $A_N$ at 1.8 GeV. 
The dashed curve is identical to the solid curve of
Fig.~7. Dotted curve and data as in Fig.~7.
}
\label{fig_8}
\end{figure}

\pagebreak

\begin{center}
  \epsfig{figure=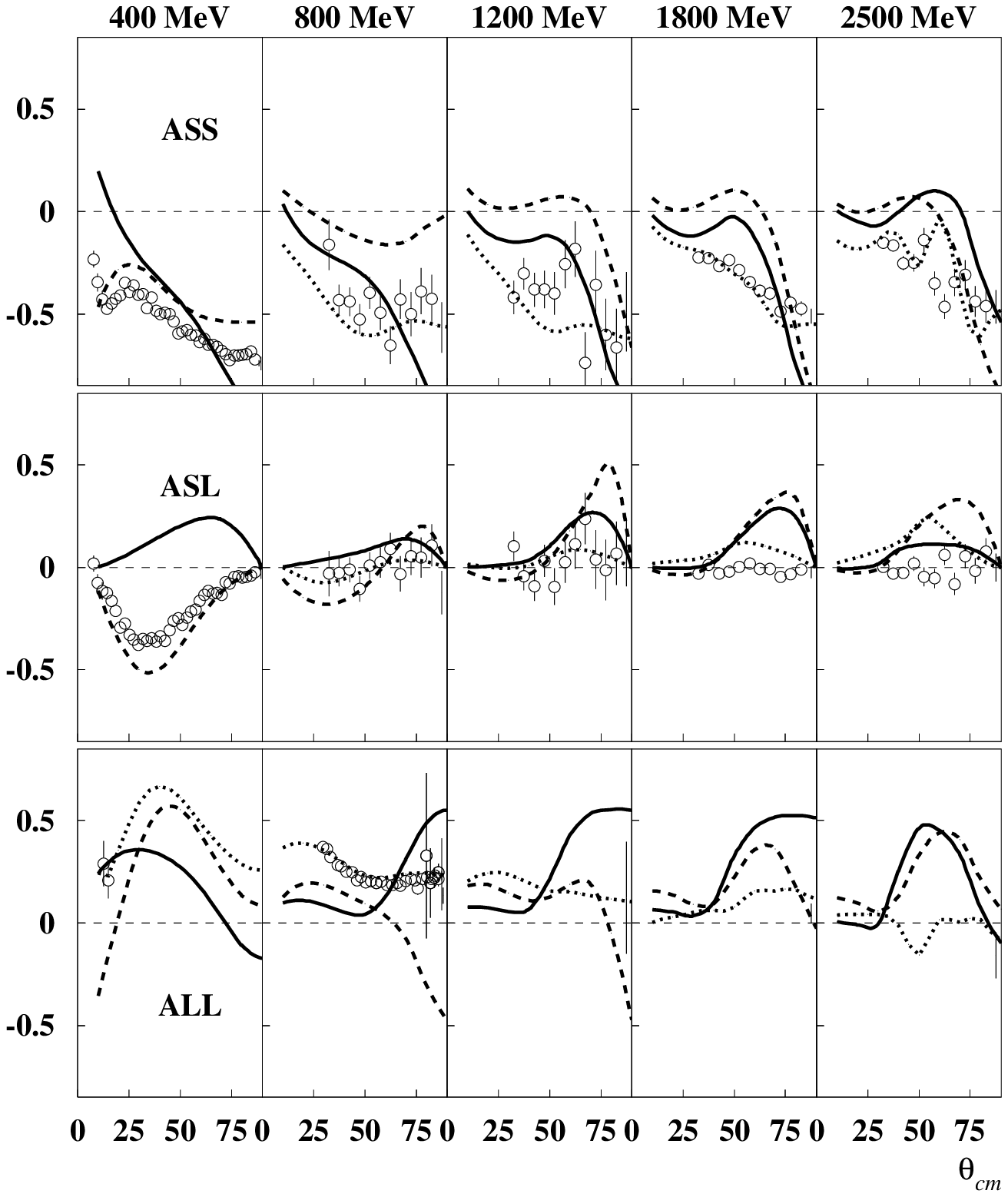,width=17cm}
  Fig.~\ref{fig_8} continued.
\end{center}


\begin{references}
\bibitem{foot1} The model presented in Sec.~II of this paper is identical
to `Model II' of appendix B of Ref.~\cite{Mac89}; see also section 7.1.2
therein.
\bibitem{foot2} Since we are anticipating modifications of the model
anyhow, the cross sections shown in Figs.~5 and 6 have been calculated
using an omega coupling constant of $g^2_\omega/4\pi = 20.6$, while
all other meson parameters have been kept at their values
given in Table~I.
\end{references}
\end{document}